\documentclass[journal,a4paper]{IEEEtran}
\usepackage{amssymb}
\usepackage{amscd}
\usepackage{amsmath}
\usepackage{epsfig}
\usepackage{amsthm}
\usepackage{graphics}
\usepackage{psfrag}
\usepackage{rotating}
\usepackage{amsmath}
\usepackage{amsfonts}
\usepackage{url}
\usepackage{color}
\usepackage{epstopdf}
\usepackage{amsthm}
\usepackage{tikz}
\usepackage{mathtools}
\usepackage{multirow}
\usepackage[final]{pdfpages}
\usepackage{enumitem}
\usepackage{multicol}
\usepackage[nolist]{acronym}
\usepackage{subcaption}
\usepackage[ruled]{algorithm2e}
\usepackage{ragged2e}

\newtheorem{remark}{Remark}

\newfont{\bbb}{msbm10 scaled 700}

\newfont{\bb}{msbm10 scaled 1100}
\newcommand{\CC}{\mbox{\bb C}}

\newcommand{\RR}{\mbox{\bb R}}

\newcommand{\EE}{\mbox{\bb E}}

\newcommand{\av}{{\bf a}}
\newcommand{\bv}{{\bf b}}

\newcommand{\fv}{{\bf f}}
\newcommand{\gv}{{\bf g}}
\newcommand{\hv}{{\bf h}}

\newcommand{\rv}{{\bf r}}
\newcommand{\sv}{{\bf s}}

\newcommand{\wv}{{\bf w}}

\newcommand{\xv}{{\bf x}}
\newcommand{\yv}{{\bf y}}

\newcommand{\Am}{{\bf A}}
\newcommand{\Bm}{{\bf B}}
\newcommand{\Cm}{{\bf C}}

\newcommand{\Fm}{{\bf F}}
\newcommand{\Gm}{{\bf G}}
\newcommand{\Hm}{{\bf H}}
\newcommand{\Id}{{\bf I}}

\newcommand{\Um}{{\bf U}}

\newcommand{\Vm}{{\bf V}}
\newcommand{\Xm}{{\bf X}}

\newcommand{\Fc}{{\cal F}}

\newcommand{\Pc}{{\cal P}}

\newcommand{\Tc}{{\cal T}}

\newcommand{\thetav}{\hbox{\boldmath$\theta$}}

\newcommand{\Psim}{\hbox{\boldmath$\Psi$}}

\newcommand{\trace}{{\hbox{tr}}}

\renewcommand{\arg}{{\hbox{arg}}}

\renewcommand{\Re}{{\rm Re}}

\newcommand{\eqdef}{\stackrel{\Delta}{=}}

\newcommand{\Pav}{P_{\rm avg}}
\newcommand{\Na}{N_{\rm a}}
\newcommand{\Nrf}{N_{\rm rf}}
\newcommand{\Ns}{N_{\rm s}}
\renewcommand{\H}{{\scriptscriptstyle\mathsf{H}}}
\newcommand{\T}{{\scriptscriptstyle\mathsf{T}}}

\begin{document}

\begin{acronym}
	\acro{AWGN}{additive white Gaussian noise}
	\acro{MIMO}{Multiple-input multiple-output}
	\acro{OTFS}{orthogonal time frequency space}
	\acro{SNR}{signal-to-noise ratio}
	\acro{mmWave}{millimeter wave}
	\acro{ML}{maximum likelihood}
	\acro{V2X}{vehicle-to-everything}
	\acro{OFDM}{orthogonal frequency division multiplexing}
	\acro{FMCW}{frequency modulated continuous wave}
	\acro{LoS}{line-of-sight}
	\acro{ISFFT}{inverse symplectic finite Fourier transform}
	\acro{SFFT}{symplectic finite Fourier transform}
	\acro{HPBW}{half-power beamwidth}
	\acro{ULA}{uniform linear array}
	\acro{CRLB}{Cram\'er-Rao Lower Bound}
	\acro{RF}{radio frequency}
	\acro{BF}{beamforming}
	\acro{RMSE}{root MSE}
	\acro{AoA}{angle of arrival}
	\acro{ISI}{inter-symbol interference}
	\acro{SI}{self-interference}
	\acro{TDD}{time division duplex}
	\acro{Tx}{transmitter}
	\acro{Rx}{receiver}
	\acro{SIC}{successive interference cancellation}
	\acro{PD}{probability of detection}
	\acro{HDA}{hybrid digital-analog}
	\acro{PSD}{power spectral density}
\end{acronym}

\title{Hybrid Digital-Analog Beamforming and MIMO Radar with OTFS Modulation}

\author{
	\IEEEauthorblockN{Lorenzo Gaudio$^{1}$, Mari Kobayashi$^{2}$, Giuseppe Caire$^{3}$, Giulio Colavolpe$^{1}$}\\
	\IEEEauthorblockA{$^{1}$University of Parma, Italy \\
		$^{2}$Technical University of Munich, Munich, Germany \\
		$^{3}$Technical University of Berlin, Germany\\ 
		Emails: lorenzo.gaudio@studenti.unipr.it, mari.kobayashi@tum.de, giulio.colavolpe@unipr.it, caire@tu-berlin.de }
	}

\maketitle

\begin{abstract}

Motivated by future automotive applications, we study some joint radar target detection and parameter estimation problems where the transmitter, equipped with a mono-static MIMO radar, wishes to detect multiple targets and then estimate their respective parameters, while simultaneously communicating information data using orthogonal time frequency space (OTFS) modulation. Assuming that the number of radio frequency chains is smaller than the number of antennas over the mmWave frequency band, we design hybrid digital-analog beamforming at the radar transmitter adapted to different operating phases. 
The first scenario considers a wide angular beam in order to perform the target detection and parameter estimation, while multicasting a common message to all possible active users. The second scenario considers narrow angular beams to send information streams individually to the already detected users and simultaneously keep tracking of their respective parameters. 
Under this setup, we propose an efficient maximum likelihood scheme combined with hybrid beamforming to jointly perform target detection and parameter estimation. 
Our numerical results demonstrate that the proposed algorithm is able to reliably detect multiple targets with a sufficient number of antennas and achieves the Cram\'er-Rao lower bound for radar parameter estimation such as delay, Doppler and angle-of-arrival (AoA).
\end{abstract}


\section{Introduction}\label{sec:Introduction}

\ac{MIMO} radar has been extensively studied and was shown to improve the resolution, i.e., the ability to distinguish multiple targets, thanks to the additional spatial dimension (see, e.g., \cite{li2008mimo}). A careful design of \ac{BF}, or power allocation along angular directions, is crucial to achieve accurate radar detection and parameter estimation performance. This is particularly relevant to automotive radar \cite{patole2017automotive} operating over \ac{mmWave} frequency bands, as the high propagation loss must be compensated by proper \ac{BF}, or more generally beam alignment both at \ac{Tx} and \ac{Rx} sides (see e.g. \cite{song2018scalable} and references therein). 
Note that, in a mono-static radar system with co-located \ac{Tx} and \ac{Rx}, transmit and receive antennas are calibrated such that their beam patterns ``look in the same direction''.  Moreover, \ac{BF} at the radar \ac{Tx} might be adaptive, depending on different operating phases (see e.g., \cite{friedlander2012transmit,niesen2019joint} and references therein). Namely, the transmitted power shall be allocated to wider angular sectors during a target detection/search phase, while narrow and distinct beams, each focused on the detected target, shall be used to minimize  ``multi-target'' interference
in a tracking phase \cite{li2008mimo,richards2014fundamentals,bar1995multitarget}. During the target detection phase, a non-trivial tradeoff appears. On one hand, a wider angular sector coverage enables to detect potentially more targets if the received backscattered power is high enough. On the other hand, a directional \ac{BF} grants higher received \ac{SNR}, at the cost of a time-consuming search over narrower angular sectors (as classical radar successively swapping adjacent regions, see, e.g., \cite{richards2014fundamentals}). Different solutions to the aforementioned problem can be found in the literature (see, e.g., \cite{niesen2019joint,buzzi2019mimo,grossi2018opportunisticRadar,fortunati2020mimoRadar}).

\begin{figure}
	\begin{subfigure}{.49\columnwidth}
		\centering
		\includegraphics[height=0.2\textheight]{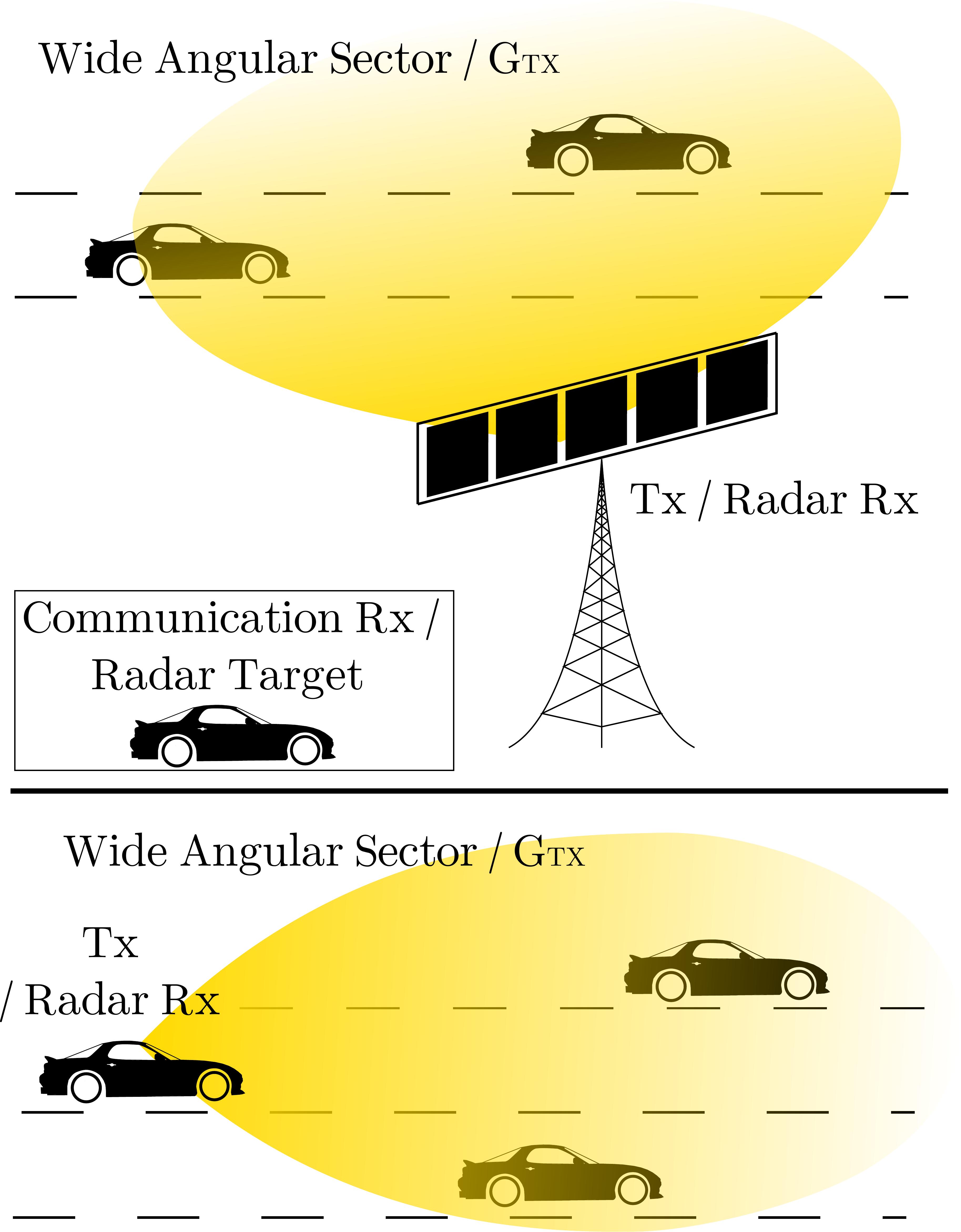}  
		\caption{Detection phase}
		\label{fig:Operational-Modes-a}
	\end{subfigure}
	\begin{subfigure}{.49\columnwidth}
		\centering
		\includegraphics[height=0.2\textheight]{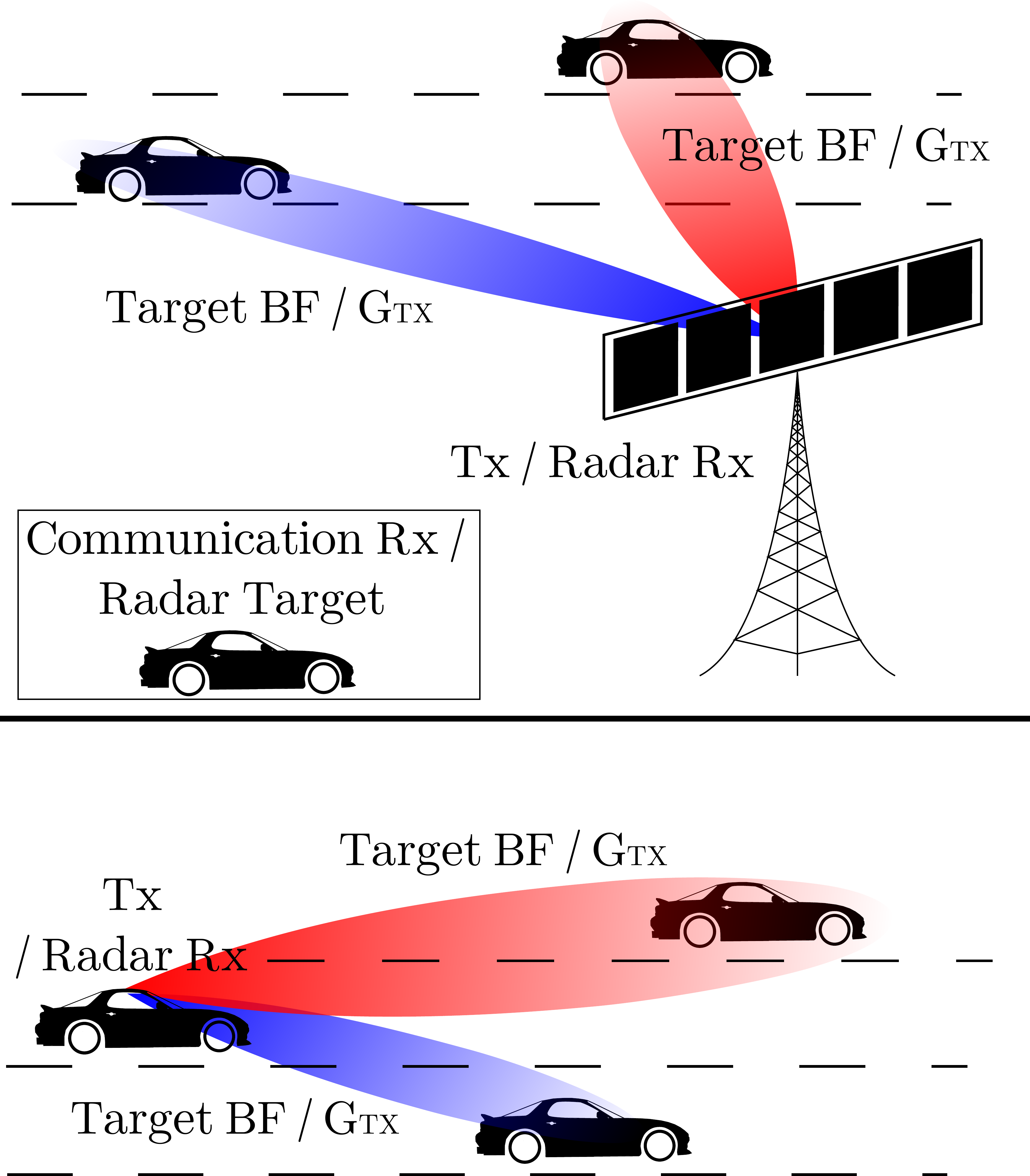}  
		\caption{Tracking phase}
		\label{fig:Operational-Modes-b}
	\end{subfigure}
	\caption{Two scenarios with two different Tx beam patterns. In (a), a Tx (a base station or a car) broadcasts a common message exploring a wide angular sector. In (b), we consider directional BF towards the detected targets. The Rx always makes use of a wide beam within the angular sector of interest.}
	\label{fig:Operational-Modes}
\end{figure}

\begin{figure}
	\centering
	\includegraphics[scale=0.12]{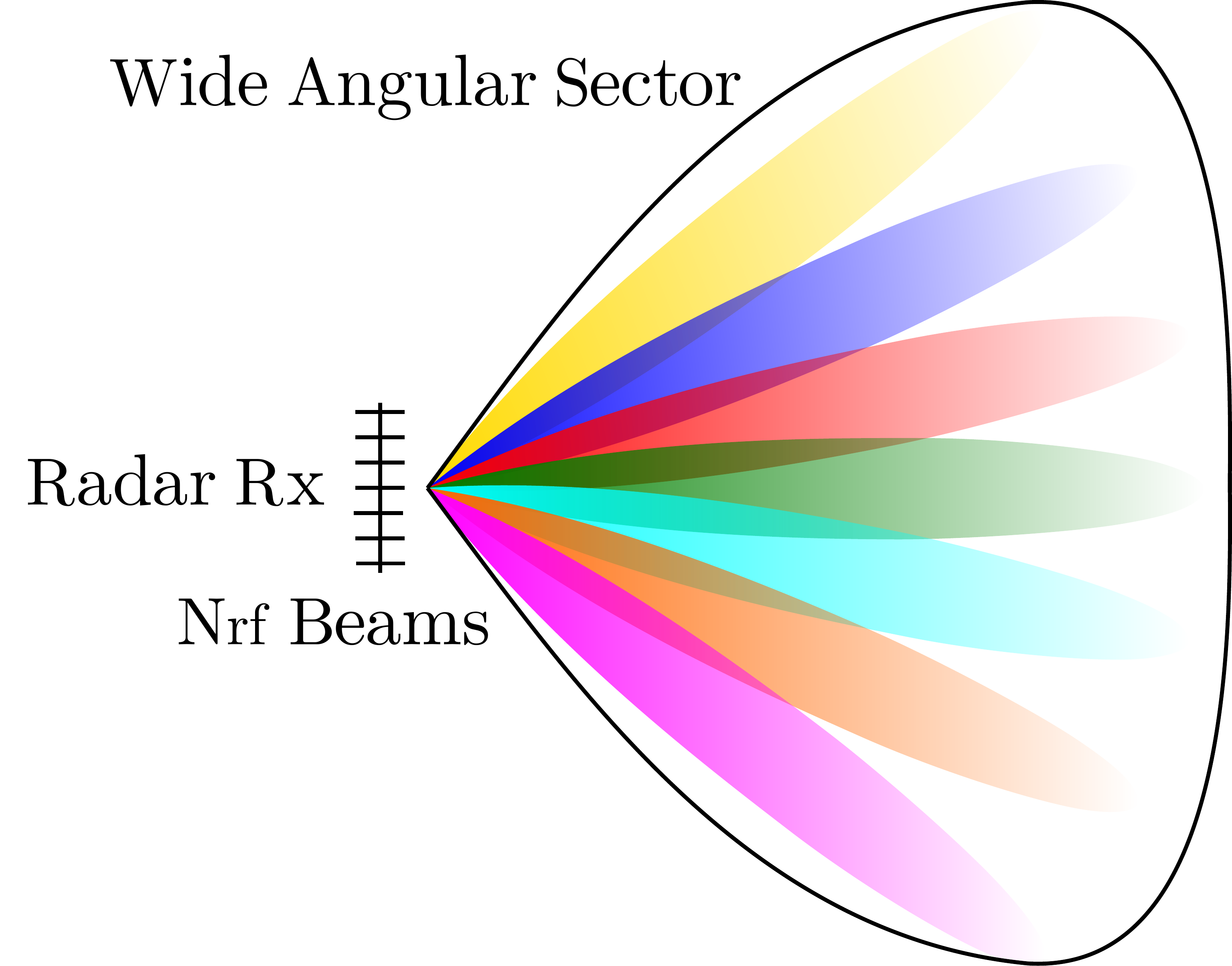}
	\caption{Beam configuration at the radar Rx for the two scenarios depicted in Fig. \ref{fig:Operational-Modes}. $\Nrf$ beams cover a wide angular sector.}
	\label{fig:Rx-Beams}
\end{figure}

As an extension of our previous works \cite{gaudio2019effectiveness,gaudio2020joint}, this paper studies the joint target detection and parameter estimation problem with a mono-static \ac{MIMO} radar adopting \ac{OTFS}, i.e., a multi-carrier modulation proposed in \cite{hadani2017orthogonal} and already studied in different \ac{MIMO} configurations (see, e.g., \cite{shen2019channel,ramachandran2018mimoOTFS}). The use of communication waveforms for radar has been motivated by the joint radar and communication paradigm, where two functions are implemented by sharing the same resources and the same waveform (see e.g. \cite{dokhanchi2019mmwave,zheng2019radar,hassanien2019dual} and references therein). Contrary to the existing works on radar sensing using \ac{OTFS} \cite{gaudio2019effectiveness,gaudio2020joint,raviteja2019orthogonal}, this paper considers \ac{MIMO} radar under a practical \ac{mmWave} system architecture such that the number of \ac{RF} chains ($\Nrf$) is much smaller than the number of antennas ($\Na$). In fact, it is difficult to implement a fully digital \ac{BF}, or, equivalently, associate one \ac{RF} chain per antenna (including A/D conversion, modulation, and amplification) in a small form factor and highly integrated technology over a large signal bandwidth. 
Therefore, focusing on \ac{mmWave} automotive applications, we consider 
\ac{HDA} \ac{BF} schemes as typically considered in the literature (see, e.g., \cite{song2019fully,chen2018doa} and references therein).
We will study two different scenarios, exploring the aforementioned \ac{BF} tradeoff. The first scenario considers a \ac{Tx} \ac{BF} design such that the beam covers a wide angular sector, to perform target detection and parameter estimation, while multicasting a common message to all possible active users (see Fig. \ref{fig:Operational-Modes-a}). The common message corresponds, for instance, to real-time traffic information that can be sent by a base station nearby or a car itself. 
Assuming that this initial communication phase is established, the second scenario considers a \ac{Tx} \ac{BF} with directed narrow beams, such that individual information streams are sent to the detected users (see Fig. \ref{fig:Operational-Modes-b}). 
Note that the Radar \ac{Rx} uses a wide beam pattern consisting of $\Nrf$ beams as illustrated in Fig. \ref{fig:Rx-Beams} to obtain a meaningful vector observation, necessary for \ac{AoA} estimation, regardless of the operating phase. This is in a sharp contrast to the hybrid beam alignment considered in the communication system where the receiver also applies \ac{BF} and obtains a scalar observation precluding the estimation of the \ac{AoA} (see, e.g., \cite{song2019fully,song2018scalable} and references therein).

Under this setup, we propose an efficient \ac{ML}-based scheme combined with \ac{HDA} \ac{BF} to jointly perform target detection and parameter estimation.
More precisely, our scheme first performs target detection and super-resolution estimation of delay, Doppler shift, and \ac{AoA} using a wide angular beam along which a single data stream is sent. Then, once the targets are detected, the subsequent tracking phase performs the parameter estimation using multiple narrow beams along which multiple data streams can be sent. Our numerical results demonstrate that the proposed scheme is able to reliably detect multiple targets while essentially achieving the \ac{CRLB} for radar parameter estimation. Furthermore we investigate various scenarios of near-far effects of targets, showing that a \ac{SIC} mechanism is able to efficiently remove the masking effect between targets located at different ranges from the radar, and we provide an in-deep analysis of the two scenarios of interest, showing their limits and advantages.

The paper is organized as follows. In Section \ref{sec:phy-model} we introduce the physical model and \ac{OTFS} modulation basics. Section \ref{sec:Joint-Detection-Param-Estimation} exploits the joint detection and parameter estimation algorithm. Numerical results are analyzed in Section \ref{sec:Numerical-Results}, and Section \ref{sec:Conclusions} concludes the paper.

We adapt the following notations.  $(\cdot)^\T$ denotes the transpose operation.  $(\cdot)^\H$ denotes the Hermitian (conjugate and transpose) operation. Operator $\left|\cdot\right|$ denotes the absolute value $\left|x\right|$ if $x\in\RR$, or the cardinality (number of elements) of a discrete set, i.e., $\left|\Fc\right|$, if $\Fc$ is a discrete set.

\section{Physical model}\label{sec:phy-model}
We consider joint radar detection and parameter estimation in a system operating over a channel bandwidth $B$ at the carrier frequency $f_c$.  We assume a \ac{Tx} equipped with a mono-static \ac{MIMO} radar with $\Na$ antennas and $\Nrf$ \ac{RF} chains, operating in full-duplex mode.\footnote{Full-duplex operations can be achieved with sufficient isolation between the transmitter and the (radar) detector and possibly interference analog pre-cancellation in order to prevent the (radar) detector saturation \cite{sabharwal2014band,wentworth2007applied,duarte2010fullduplexWireless}.}
The radar \ac{Rx} (which is colocated with the \ac{Tx}) processes the backscattered signal to identify the presence of targets within the beam, while estimating parameters of interest such as range, velocity, and \ac{AoA}. A point target model is taken into account, such that each target can be represented through its \ac{LoS} path only \cite{kumari2018ieee,nguyen2017delay,grossi2018opportunisticRadar}. By letting $\phi\in [-\frac{\pi}{2}, \frac{\pi}{2}]$ be the steering angle,  by considering an antenna array with $\lambda/2$ spacing ($\lambda$ is the wavelength), the \ac{Tx} and \ac{Rx} arrays are given by $\av(\phi)$ and $\bv(\phi)$ respectively, where $\av(\phi)=(a_1(\phi), \dots, a_{N_a}(\phi))^\T\in \CC^{N_a}$, denotes the uniform linear array response vector of the radar \ac{Rx} with 
\begin{align}\label{eq:ULA}
	a_n(\phi)&= e^{j(n-1)\pi \sin(\phi)},\;\; n=1,\dots, N_a,
\end{align}  
and $b_n(\phi)=a_n(\phi)$. Under the mono-static radar assumption, i.e., same angle $\phi$ at radar \ac{Tx} and \ac{Rx}, vectors $\av$ and $\bv$ result to be equal.
The channel is modeled as a $P$-tap time-frequency selective channel of dimension $\Na\times\Na$ given by \cite{vitetta2013wireless}
\begin{align}\label{eq:Channel}
	\Hm (t, \tau) =  \sum_{p=0}^{P-1} h_p \bv(\phi_p) \av^\H (\phi_p)\delta(\tau-\tau_p)  e^{j2\pi \nu_p t}\,,
\end{align}
where $P$ is the number of targets, $h_p$ is a complex channel gain including the pathloss, $\nu_p = \frac{2v_p f_c}{c}$ is the round-trip Doppler shift, $\tau_p=\frac{2r_p}{c}$ is the round-trip delay, and $\phi_p$ denotes the \ac{AoA}, each corresponding to the $p$-th target, respectively.
\subsection{OTFS Input Output Relation}\label{subsec:OTFS-Input-Output}
We consider \ac{OTFS} with $M$ subcarriers of bandwidth $\Delta f$ each, such that the total bandwidth is given by $B=M\Delta f$. We let $T$ denote the symbol time, yielding the \ac{OTFS} frame duration of $NT$, with $N$ number of symbols in time. We let $T\Delta f=1$, as typically considered in the \ac{OTFS} literature \cite{gaudio2019effectiveness,hadani2017orthogonal,raviteja2018interference}. 
In order to consider the aforementioned different operational modes, we let $\Ns$ denote the number of data streams to be sent per time-frequency domain such that $\Ns=1$ corresponds to the multicasting of a single data stream and $\Ns\leq \Nrf$ corresponds to the broadcasting of individual data streams.
Following the standard derivation of the input-output relation of \ac{OTFS} (see, e.g., \cite{hadani2017orthogonal,gaudio2019effectiveness}), $\Ns$-dimensional data symbols $\left\{\xv_{k,l}\right\}$, for $k=0,\dots,N-1$, $l=0,\dots,M-1$, belonging to any constellation, are arranged in an $N\times M$ two-dimensional grid referred to as the Doppler-delay domain, i.e., $\Gamma=\left\{\left({k}/{NT},{l}/{M\Delta f}\right)\right\}$. 
The \ac{Tx} first applies the \ac{ISFFT} to convert data symbols $\{\xv_{k,l}\}$ into a $\Ns\times1$ block $\{\Xm[n, m]\}$ in the time-frequency domain
\begin{equation}\label{eq:x-to-X}
\Xm[n,m]=\sum_{k=0}^{N-1}\sum_{l=0}^{M-1}\xv_{k,l}e^{j2\pi\left(\frac{nk}{N}-\frac{ml}{M}\right)}, 
\end{equation}
for $n = 0, \ldots, N-1$, $m = 0, \ldots, M-1$,  
satisfying the average power constraint $\EE[\Xm[n,m]^\H\Xm[n,m]]=\Pav/(NM \Na) \Id_{\Na}$, where $\Id_{\Ns}$ denotes an identity matrix of dimension $\Ns$. 
After assigning $\Ns$ streams to $\Nrf$ RF chains through a mapping matrix $\Vm\in \CC^{\Nrf \times \Ns}$ , the \ac{Tx} generates the $\Nrf$-dimensional continuous-time signal
\begin{align}\label{eq:Tx-Signal-General}
\sv(t)  = \Vm \sum_{n=0}^{N-1} \sum_{m=0}^{M-1} \Xm[n, m] g_{\rm tx}(t-nT) e^{j 2\pi m \Delta f(t-nT)}.
\end{align}
Since the number of RF chains is typically much smaller than the number of antennas, different types of \ac{HDA} architectures between \ac{RF} chains and antennas have been considered in the literature (see e.g. \cite{song2019fully}). In this paper, we focus on the fully-connected \ac{HDA} scheme of \cite{song2019fully}. 
For any \ac{HDA} architecture, the transmitter applies the hybrid \ac{BF} matrix denoted by $\Fm\in \CC^{\Na\times \Nrf}$ that captures both baseband and 
\ac{RF} analog \ac{BF} (see \cite{friedlander2012transmit,fortunati2020mimoRadar}), while the receiver sees the received signal of a reduced dimension through a projection matrix denoted by 
$\Um\in \CC^{\Nrf\times \Na}$. By imposing $\trace(\Fm\Vm\Vm^\H\Fm^\H)=\Na$, the total power constraint $\Pav$ is satisfied. 

By transmitting the signal \eqref{eq:Tx-Signal-General} over the channel \eqref{eq:Channel}, the $\Nrf$-dimensional continuous-time received signal is given by 
\begin{align}\label{eq:Received-Signal-First}
	\rv(t) = \sum_{p=0}^{P-1} h_p {\Um} \bv(\phi_p) \av^\H(\phi_p)\Fm \sv(t-\tau_p) e^{j2\pi \nu_pt}\,.
\end{align}

The output of the \ac{Rx} filter-bank adopting a generic receive shaping pulse $g_{\rm rx}(t)$ is given in \eqref{eq:y-received}, shown at the top of next page.
\begin{figure*}
	\begin{align}\label{eq:y-received}
	\yv(t,f) &= \int  \rv(t')g^*_{\rm rx} (t'-t) e^{-j2\pi ft'} dt'= \int_{t'} g^*_{\rm rx} (t'-t) \sum_{p=0}^{P-1} h_p {\Um}\bv(\phi_p) \av^\H(\phi_p)\Fm\sv(t'-\tau_p) e^{j2\pi \nu_pt'} e^{-j2\pi ft'} dt'\nonumber\\
	&= 
	\sum_{p,n',m'}h_p {\Um}\bv(\phi_p)\av^\H(\phi_p) \Fm \Vm \Xm[n', m']\int_{t'} g^*_{\rm rx} (t'-t)  g_{\rm tx}(t'-\tau_p-n'T) e^{j 2\pi m' \Delta f(t'-\tau_p-n'T)} e^{j2\pi (\nu_p-f) t'} dt'
	\end{align} 
\end{figure*}
By sampling at $t=nT$ and $f=m\Delta f$, we obtain
\begin{align}
\yv[n, m] &=\yv(t, f)|_{t=n T}^{f=m\Delta f}=\sum_{n'=0}^{N-1} \sum_{m'=0}^{M-1}\hv_{n,m}[n', m']\,,
\end{align}
where the time-frequency domain input-output relation $\hv_{n,m}[n', m']$ is given in \eqref{eq:time-frequency-h}, shown at the top of next page.
\begin{figure*}
	\vspace{-0.5cm}
	\begin{align}\label{eq:time-frequency-h}
	\hv_{n,m}[n', m']= \sum_{p=0}^{P-1} h'_p {\Um}\bv(\phi_p) \av^\H(\phi_p)\Fm  \Vm \Xm[n', m']C_{g_{\rm tx},g_{\rm rx}}((n-n')T -\tau_p, (m-m') \Delta f-\nu_p) e^{j 2\pi n' T \nu_p} e^{-j2\pi m \Delta f  \tau_p}
	\end{align}
\end{figure*}
Notice that we defined the cross ambiguity function $C_{u, v}(\tau, \nu) \eqdef \int_{-\infty}^{\infty} u(s) v^*(s-\tau) e^{-j 2\pi \nu s} ds$ as in \cite{matz2013time}, let $h'_p = h_p e^{j 2\pi \nu_p \tau_p}$, and imposed the term $e^{-j2\pi mn'\Delta f T}=1$, $\forall n',m$, under the hypothesis $T\Delta f=1$.
Since each $X_i[n, m]$ is generated via \ac{ISFFT}, the received signal in the delay-Doppler domain is obtained by the application of the \ac{SFFT}
\begin{align}
\yv[k,l]=\sum_{n,m}\frac{\yv[n,m]}{NM}e^{j2\pi\left(\frac{ml}{M}-\frac{nk}{N}\right)}=\sum_{k', l'}\gv_{k,k'}\left[l,l'\right],
\end{align}
where the \ac{ISI} coefficient of the Doppler-delay pair $\left[k',l'\right]$ seen by sample $\left[k,l\right]$ is given by
\begin{equation}
\gv_{k,k'}\left[l,l'\right]=\sum_p h_p' {\Um}\bv(\phi_p)\av^\H(\phi_p)\Fm\Vm\xv_{k',l'}\Psi^p_{k, k'}[l,l']\,,
\end{equation}
with $\Psi^p_{k, k'}[ l, l']$ defined in \eqref{eq:Psi-Mat}, shown at the top of next page.
\begin{figure*}
	\vspace{-0.5cm}
	\begin{align}\label{eq:Psi-Mat}
	\Psi^p_{k, k'}[ l, l'] =\!\!\! \sum_{n, n', m, m'}\!\!\! \frac{C_{g_{\rm rx}, g_{\rm tx}}((n-n')T -\tau_p, (m-m') \Delta f-\nu_p)}{NM}  e^{j 2\pi n' T \nu_p} e^{-j2\pi m \Delta f  \tau_p}e^{j 2\pi \left(\frac{n'k'}{N}- \frac{m'l'}{M}\right)}e^{-j 2\pi\left(\frac{nk}{N}-\frac{ml}{M}\right)}
	\end{align}
	\noindent\rule{\textwidth}{0.4pt}
\end{figure*}
A simplified version of $\Psi^p_{k, k'}[ l, l']$ obtained by approximating the cross ambiguity function can be found in \cite{gaudio2019effectiveness}.

\subsection{Beamforming matrices}
The design of \ac{BF} matrix $\Fm$ at the radar \ac{Tx} depends on the operating phase. During the detection phase depicted in Fig.~\ref{fig:Operational-Modes-a}, we choose $\Fm$ for a given angular sector such that both \ac{Tx} and \ac{Rx} are aligned towards the same wide angular sector. 
Following \cite[Section III.C]{friedlander2012transmit}, we construct $\Fm\in \CC^{\Na\times \Nrf}$ to cover a wide angular sector $[-\theta, \theta]$ as follows. 
By representing this angular sector by a discrete set of $\Nrf$ angles, denoted by $\Theta=\left\{ \pm \left({\theta}/{(2\Nrf)} + k{\theta}/{\Nrf} \right)\right\}$, for $k=0,\dots,\Nrf/2-1$, we construct each column of $\Fm = [\fv_1, \dots, \fv_{\Nrf}]$ as 
\begin{align}\label{eq:columnF}
\fv_i = \frac{\av(\theta_i)}{|\av(\theta_i)|}, \;\; i=1, \dots, \Nrf\,,
\end{align}
with a suitable normalization, where $\av(\theta_i)$ is defined in \eqref{eq:ULA}. 

During the target tracking phase, we form multiple narrow beams corresponding to the estimated \ac{AoA} of the detected targets. This is illustrated with red and blue beams, corresponding to two different \ac{AoA}, in Fig.~\ref{fig:Operational-Modes-b}. Assuming that $P$ targets are detected and their respective AoA are estimated, we construct $\Fm$ by replacing $\theta_i$ by $\hat{\phi}_p$  for the first $P$ columns in \eqref{eq:columnF} \cite[Section III.B]{friedlander2012transmit}.

Contrary to the transmit beamforming matrix, the reduction matrix $\Um$ at the radar \ac{Rx} remains the same for both detection and tracking phases. Namely, we set 
 $\Um = \Fm^\H$,  where each column is given in \eqref{eq:columnF}. This is illustrated in Fig.~\ref{fig:Rx-Beams}.  
This choice of an isotropic receive beam enables to obtain a multi-dimensional signal for \ac{AoA} estimation in both detection and tracking phases.

\section{Joint Detection and Parameters Estimation}\label{sec:Joint-Detection-Param-Estimation}
We wish to estimate the set of four parameters $\thetav = \{h'_p, \phi_p, \tau_p, \nu_p\}\in \Tc^P$, with $\Tc=  \CC \times \RR \times \RR \times \RR$. By defining
\begin{align}\label{eq:G-Matrix}
	\Gm_{p}(\tau_p, \nu_p,\phi_p)\triangleq\left({\Um}\bv\left(\phi_p\right)\av^\H(\phi_p)\Fm \Vm\right)\otimes\Psim^p\,,
\end{align} 
where $\otimes$ is the Kronecker product,\footnote{Note that $\Am^{X\times Y}\otimes \Bm^{Z\times K}=\Cm^{XZ\times YK}$.} as the ${\Nrf} NM\times \Ns NM$ matrix obtained by multiplying $\Psim^p$ by a different coefficient of $\left({\Um}\bv\left(\phi_p\right)\av^\H(\phi_p)\Fm \Vm\right)$. Thus, by stacking $\Xm$ into a $\Ns NM$-dimensional vector $\xv$ and defining an output vector $\yv$ of dimension $NM{\Nrf}\times 1$,
the received signal in the presence of noise is given by
\begin{align}\label{eq:Received-signal}
\yv = \sum_{p=0}^{P-1} \left[h'_p\Gm_p(\tau_p, \nu_p,\phi_p)\right]\xv  + \wv\,,
\end{align}
where $\wv$ denotes the \ac{AWGN} vector with independent and identically distributed entries of zero mean and variance $\sigma_w^2$. The problem reduces to detect $P$ targets and estimate the $4P$ associated parameters (complex channel coefficient, Doppler, delay, and angle) from the ${\Nrf}MN$-dimensional received signal.  To this end, we define the \ac{ML} function as 
\begin{align}\label{eq:497}
l(\yv| \thetav, \xv) & = \left|\yv - \sum_p h'_p \Gm_p \xv  \right|^2, 
\end{align}
where we use the short hand notation $\Gm_p\triangleq\Gm(\tau_p, \nu_p,\phi_p)$.  The \ac{ML} solution is given by 
\begin{align}
\hat{\thetav} = \arg\min_{\thetav \in \Tc^P}  l(\yv| \thetav, \xv).
\end{align}
For a fixed set of $\{ \phi_p, \tau_p, \nu_p\}$, the \ac{ML} estimator of $\{h'_p\}$ is given by solving the following set of equations 
\begin{align}\label{eq:459}
\xv^\H\Gm^\H_p\left(\sum_{q=0}^{P-1} h'_q  \Gm_q\right) \xv &=  \xv^\H  \Gm^\H_p \yv, \;\;\; p=0,\dots, P-1.  
\end{align}
By plugging \eqref{eq:459} into \eqref{eq:497}, it readily follows that minimizing $ l(\yv| \thetav, \xv)$ reduces to maximize the following function
\begin{align}
l_2(\yv| \thetav, \xv) &= \sum_p h'_p  \yv^\H\Gm_p \xv 
\nonumber\\
&= \sum_p S_p(\tau_p, \nu_p, \phi_p) - I_p( \{h'_q\}_{q\neq p}, \thetav)\,,
\end{align}
where $S_p(\tau_p, \nu_p, \phi_p)$ and  $I_p( \{h'_q\}_{q\neq p},\thetav)$ ($S_p$ and $I_p$ in short hand notation) denote the useful signal and the interference term for target $p$, given respectively by
\begin{align}\label{eq:Sp}
S_p  =  \frac{| \yv^\H\Gm_p \xv|^2}{| \Gm_p \xv|^2}\,,
\end{align}
\begin{align}
I_p =\frac{\left(\yv^\H\Gm_p \xv\right) \xv^\H  \left(\Gm^\H_p\sum_{q\neq p}h'_q  \Gm_q\right)  \xv}{| \Gm_p \xv|^2}\,.
\end{align}

Notice that we have $I_p=0$ if there is only one target.

\subsection{Successive Interference Cancellation (SIC) and Joint Target Detection and Parameters Estimation Algorithm}

\SetKwRepeat{Repeat}{Repeat}{\hspace{-0.3cm}Until}
\SetKwFor{For}{\hspace{-0.1cm}For}{do}{\hspace{-0.3cm}End}
\setlength{\algomargin}{0.1cm}
\begin{algorithm}
	\SetAlgoLined
	\justifying
	\noindent
	\KwResult{Target detection and radar parameter estimation.}
	\noindent\textbf{Initialization:} Set $\yv'=\yv$; Detected targets $N_{\mathrm{dt}}=0$\;
	\noindent\Repeat{Stopping Criterion}
	{
		{\noindent\textbf{1) Detection / (AOA, Doppler, Delay) Coarse Estimation:}
			Given $\yv'$, search a possible set of targets
			\begin{align}\label{eq:3D-Search}
			\Pc=&\left\{\max_{\left(\tau,\nu\right)}S_p(\tau, \nu, \phi)>T_r\right\},\\
			&\mathrm{s.t.}\ \Big\{\forall\left(\tau,\nu,\phi\right)\in\Gamma\times\Omega\Big\}\,,\nonumber
			\end{align}
			where $T_r$ is the detection threshold, to be properly optimized, $\Gamma$ is the Doppler-delay grid described in \ref{subsec:OTFS-Input-Output} and $\Omega$ is defined as a discretized set of angles. The $p$-th target is associated to a coarse estimation $(\hat{\phi}_p,\hat{\tau}_p,\hat{\nu}_p)$, such that $S_p(\hat{\phi}_p,\hat{\tau}_p,\hat{\nu}_p)$ is above the threshold and is a local maximum}\;
		\noindent\textbf{2) Super-Resolution Parameter Estimation:}\\
		\textbf{\textit{2.1) Fine AOA}:} 
		\vspace{-0.1cm}
		\begin{align}\label{eq:Phi-Fine}
		\hat{\phi}_p=\arg\max_{\phi}S_p(\hat{\tau}_p,\hat{\nu}_p,\phi)\,,\ \ p=1,\dots,\left|\Pc\right|.
		\end{align} 
		\textbf{\textit{2.2) Fine Doppler-delay Estimation}:}\\
		\textit{Initialization:} Iteration $i=0$, initialize $\hat{h}'_p[0] = 0$.\\
		\For{Iteration $i=1,2,\dots$}{
			$\bullet$ \textbf{\textit{Delay and Doppler update}}: Find the estimates $\hat{\tau}_p[i],\hat{\nu}_p[i]$ 
			by solving the two-dimensional maximization 
			\begin{align}\label{eq:Argmax-Sp-Ip}
			(\hat{\tau}_p[i],\hat{\nu}_p[i]) & = \arg\max_{\left(\tau,\nu\right)} \Big \{ S_p - I_p \Big \}\,,
			\end{align}
			with $S_p$ and  $I_p$ computed for $(\hat{h}'_p[i],\tau,\nu, \hat{\phi}_p[i])$\;
			$\bullet$ \textbf{\textit{Complex channel coefficients update}}: Solve the linear system (\ref{eq:459}) using channel matrices $\Gm_p$ with parameters $(\hat{h}'_p[i],\hat{\tau}_p[i],\hat{\nu}_p[i], \hat{\phi}_p)$, and let the solution
			be denoted by $\hat{h}'_p[i]$\;
		}
		\noindent\textbf{3) Re-Fine AOA:} Compute \eqref{eq:Phi-Fine} using the refined estimation $\left(\hat{\tau}_p,\hat{\nu}_p\right)$ obtained in \eqref{eq:Argmax-Sp-Ip}\;
		\noindent\textbf{4) SIC:} Compute
		\vspace{-0.1cm}
		\begin{align}\label{eq:SIC-Cancellation}
		y'=y-\sum_{p=0}^{\left|\Pc\right|}\left[\hat{h}'_p\Gm_p(\hat{\tau}_p,\hat{\nu}_p,\hat{\phi}_p)\right]\xv\,,
		\end{align}
		increase targets counter $N_{\mathrm{dt}}=N_{\mathrm{dt}}+\left|\Pc\right|$\;
	}
	\caption{\textit{Joint Detection and Radar Parameters Estimation}}
	\label{alg:Joint-Det-Par-Est}
\end{algorithm}

The use of OTFS for radar tasks introduces some limitations. In particular, OTFS with $T\Delta f=1$ yields a cross ambiguity function $C_{\rm g_{tx}, g_{rx}}(\tau, \nu)$ that incurs significant side-lobes in the Doppler-delay domain. 
As a result, our simulations show that the magnitude of the useful signal, i.e., $\max_{(\tau,\nu)}S_p(\tau, \nu, \phi)$,  has a main lobe around the angle $\phi_p$ of target $p$ and non-negligible side-lobes in the angular domain. 
Since the signal magnitude strictly depends on the received backscattered power, 
the sidelobes of a strong target, closer to the radar, may completely mask 
the main lobe of weaker targets, far from the radar. Therefore, situations of near-far effect among the targets, causing large power imbalance in the backscatter waves,  must be handled explicitly by some additional signal processing. 
This motivates us to incorporate \ac{SIC} in our ML-based target detection. 
Given the received signal in \eqref{eq:Received-signal}, once a strong target is detected and its radar parameters are estimated, its contribution, and thus the masking effect, can be removed from the received signal (see \eqref{eq:SIC-Cancellation} in Algorithm \ref{alg:Joint-Det-Par-Est}). This process can be run iteratively to cancel the contributions of new detected targets, until a given condition or stopping criteria is satisfied (e.g., a target is found in an angular sector already explored, or the magnitude of the useful signal goes below a certain value). Algorithm \ref{alg:Joint-Det-Par-Est} describes the steps to perform joint detection and radar parameters estimation. Some remarks on Algorithm 1 are in order:
\begin{remark}[Target Detection] 
	Equation \eqref{eq:3D-Search} presents a threshold test requiring the search over a three dimensional grids composed of $|\Omega|$ slices of the $N\times M$ Doppler-delay grid. 
	In order to keep the complexity low, we consider the Doppler-delay grid $\Gamma$ defined in Section \ref{subsec:OTFS-Input-Output} and a coarse $\Omega$.\footnote{For instance, with an angular sector covering of $60$ degrees divided in $4$ equally spaced parts, the set of angles results to be (supposing the center of the beam to be at $0$ degree) $\Omega=\{[-30,-15],[-15, 0],[0,15],[15,30]\}$.} Even if this assumption is rather restrictive, it provides a computationally feasible and fast coarse estimation (step 1 of Algorithm \ref{alg:Joint-Det-Par-Est}), to be used as a baseline for the successive super-resolution \ac{ML}-based parameter estimation (step 2 of Algorithm \ref{alg:Joint-Det-Par-Est}).
\end{remark}
\begin{remark}[Fine \ac{AoA} Estimation]
	Since $S_p$ is a convex function in $\phi$ for a fixed pair $(\tau_p,\nu_p)$, the result of \eqref{eq:Phi-Fine} can be exactly computed using common convex solvers. Therefore, the angle can be estimated with super-resolution far beyond the discrete grid $\Omega$.
\end{remark}
\begin{remark}
	The magnitude of $S_p$ strictly depends on the target range (and pathloss). Thus, in order to keep a fixed threshold $T_r$ for all iterations, the argument $\max_{(\tau,\nu)}S_p(\tau, \nu, \phi)$ has to be normalized at each iteration, for instance, w.r.t. its mean computed over all possible angles.  
\end{remark}

\subsection{Reduced-Complexity Parameter Estimation}\label{subsec:Suboptimal-Solution}
In the target tracking phase, the matrix $\Gm_p$ in \eqref{eq:G-Matrix} shall be updated dynamically as the \ac{BF} matrix $\Fm$ of dimension $\Na\times \Nrf$ and the channel matrix $\Psim^p$ of $NM\times NM$ change in time. 
The following solution can be adopted in order to reduce the computational complexity related to the dynamically changing matrices. Namely, we compute $\Gm_p$ for target $p$ by selecting only the column of $\Fm$ corresponding to this target already detected in Step 1 of Algorithm \ref{alg:Joint-Det-Par-Est}. Assuming that targets are located with different ranges from the radar, this low-complexity method does not affect the parameter estimation performance. If there are a few targets located with a similar range from the radar, they can be grouped together within the same narrow beam.

\subsection{Cram\'er-Rao Lower Bound (CRLB)}\label{sec:Cramer-Rao-Bound}
\newcommand{\Pbar}{ \bar{\Psi}}
We consider the \ac{CRLB} as a theoretical benchmark. In order to estimate a complex channel coefficient, we let $A_p = |h'_p|$ and $\psi_p = \angle(h'_p)$ denote the amplitude and the phase of $h'_p$, respectively. Thus, $5P$ real variables have to be estimated, i.e., $	\thetav =\{A_p, \psi_p, \tau_p, \nu_p,\phi_p\}$. We form the $5P\times 5P$ Fisher information matrix whose $(i,j)$ element is
\begin{align}
[\Id(\thetav) ]_{i,j} =
\frac{2}{N_0} \Re\left\{ \sum_{n, m, t} \left[\frac{\partial s_p^{[n ,m, t]}}{\partial \theta_i}\right]^* \left[\frac{\partial s_q^{[n, m,t]}}{\partial \theta_j}\right]\right\}\,, 
\end{align}
where $p = [i]_P$, $q= [j]_P$, and
\begin{align}
s_p^{[n,m,t]} &= A_p e^{j \psi_p} b_t(\phi_p)a_t^*(\phi_p)f_t \sum_{k,l}\Psi^p_{n, k}[ m, l] x_{k, l}\,,
\end{align}
where $(n,m,t)$ denote time, subcarrier, and antenna, respectively, while $f_t$ is the $t$-th entry of the \ac{BF} vector of any \ac{RF} chain.\footnote{Here we assume that, given a proper \ac{BF} design, the beam patterns directed to different targets do not interfere. Hence, we completely neglect beam interference, and only a \ac{BF} vector entry $f_t$ appears at a time.} Note that, even if not explicitly indicated for the sake of simplicity, the summations w.r.t. $k$ and $l$ extend from $0$ to $N-1$ and $M-1$, respectively, as in all previous analysis. The desired \ac{CRLB} is obtained taking the diagonal elements of the inverse Fisher information matrix, filled with the corresponding derivatives. 

\section{Numerical Results}\label{sec:Numerical-Results}
\renewcommand{\arraystretch}{1.3}
\begin{table}
	\caption{System parameters}
	\centering
	\begin{tabular}{|c|c|}
		\hline
		$N=6$ & $M=512$ \\ \hline
		$f_c=24.25$ [GHz] & $B=150$ [MHz] \\ \hline
		$v_{\mathrm{res}}\simeq440$ [km/h] & $r_{\mathrm{res}}\simeq1$ [m]\\ \hline
		$v_{\mathrm{max}}=N\cdot v_{\mathrm{res}}$ & $r_{\mathrm{max}}=M\cdot r_{\mathrm{res}}$ \\ \hline
		$\Pav=40$ [mW] & $\sigma_{\mathrm{rcs}}=1$ [m$^2$] \\ \hline
		Noise Figure $=3$ [dB] & Noise PSD = $2\cdot10^{-21}$ [W/Hz] \\ \hline
		$\Na=16,32,64,128$ & $\Nrf=8$ \\ \hline
	\end{tabular}
	\label{tab:System-Parameters}
\end{table}

We set the number of \ac{RF} chains to $\Nrf=8$, such that a single equipment is able to jointly track and communicate to $\Nrf$ separately targets (or groups of targets), while $\Nrf\ll\Na$. Table \ref{tab:System-Parameters} provides all the system parameters. 

In our simulations, we rely on the following assumptions:
\begin{itemize}
	\item Given the choice of a \ac{mmWave} communication, we assume a single \ac{LoS} path between the \ac{Tx} and the radar target \cite{kumari2018ieee,nguyen2017delay,grossi2018opportunisticRadar}. This is motivated by the fact that any possible scattering component different from the \ac{LoS} generally brings much lower power, given by additional reflections of the echo signal.
	\item Any backscattered power to the radar \ac{Rx} is considered as a possible target. The objective is to sense the surrounding environment, and the differentiation between active targets and obstacles is a post-processing decision. Clearly, in a second phase, communication is established only towards active targets. 
	\item 
	We consider the complete blockage of the signal propagation to the first object hit. This assumption is completely fulfilled in \ac{mmWave} communication scenarios.\footnote{Note that the proposed algorithm could be able to correctly distinguish more targets sharing the same angular direction, if separated in at least one other domain (Doppler or delay) \cite{gaudio2020joint}.} 
\end{itemize}
Note that the aforementioned assumption are shared by many works in literature (see, e.g., \cite{grossi2018opportunisticRadar} and references therein).

The radar two-way pathloss is defined as \cite[Chapter 2]{richards2014fundamentals}
\begin{equation}
\mathrm{PL}=\frac{(4\pi)^3r^{4}}{\lambda^2}\,,
\end{equation}
and the definition of the radar \ac{SNR} becomes 
\begin{equation}\label{eq:SNR-Formula}
\mathrm{SNR}=\frac{\lambda^2\sigma_{\mathrm{rcs}}G_\mathrm{Tx}G_\mathrm{Rx}}{\left(4\pi\right)^3r^4}\frac{\Pav}{\sigma_w^2}\,,
\end{equation}
where $\lambda=c/f_c$ is the wavelength, $c$ is the speed of light, $\sigma_{\mathrm{rcs}}$ is the radar cross-section of the target in $\mathrm{m}^2$, $G_\mathrm{Tx}$ and $G_\mathrm{Rx}$ are the antenna gains at the \ac{Tx} and \ac{Rx} respectively, $r$ is the distance between \ac{Tx} and \ac{Rx}, and $\sigma_w^2$ is the variance of the \ac{AWGN} noise with noise \ac{PSD} of $2\cdot10^{-21}$ [W/Hz]. We choose $\sigma_{\mathrm{rcs}}=1$ [m$^2$], while different choices can be found in literature \cite{suzuki2000measurement,kamel2017RCSmodeling}. Note that, while $G_\mathrm{Tx}$ can change with the operational mode, $G_\mathrm{Rx}$ is keep constant (within the angular sector of interest) in order to allow isotropic reception, as already explained. Information about antenna gains, beam patterns, two-way (\ac{Tx} and \ac{Rx}) beamwidth, and more antenna basics (also for mono-static radars) can be found, for instance, in \cite{visser2006array,friedlander2012transmit}. The detection threshold $T_r$ in Algorithm \ref{alg:Joint-Det-Par-Est} has been numerical evaluated in order to have a false alarm probability of $10^{-4}$ (as done, e.g., in \cite{grossi2018opportunisticRadar}).

While two distinct targets in the angle domain can be identified if the angular resolution meets some conditions (depending on the number of antennas, the angular distance between the two targets, and the antenna array properties) \cite{richards2014fundamentals}. The velocity and the range resolution is determined by the system parameters in Table \ref{tab:System-Parameters} and given by
\begin{equation}\label{eq:Radar-Resolution}
v_{\mathrm{res}}=\frac{Bc}{2NMf_c}\ \mathrm{[m/s]}\,,\:\:\:r_{\mathrm{res}}=\frac{c}{2B}\ \mathrm{[m]}\,.
\end{equation}
In order to get a reasonable range resolution, e.g., $<1$ [m], a large bandwidth has to be considered.\footnote{Note that a tradeoff appears. Larger bandwidths mean more precise resolution, but lower theoretical maximum range (with the same $N\times M$ grid). We remark that our algorithm is completely independent of these choices.} Since the velocity resolution is directly proportional to $B$, for a fixed $f_c$, the only way to obtain lower values is to increase the block size $NM$, leading to a remarkable increase in computational complexity, which could be not affordable. For this reason, we set the system parameters by focusing on a reasonable range resolution (and theoretical maximum range) under a feasible computational complexity. Note that the maximum range could not be achieved if the backscattered power is below the noise floor. However, the chosen system setup leads to an unavoidable very large velocity resolution. Under the aforementioned assumptions, taken at the beginning of Section \ref{sec:Numerical-Results}, the problem of targets identifiability appears only in the angular domain. However, this only happens at \ac{mmWave}, thus, range and velocity resolutions are reported here for completeness, since the proposed scheme could target lower frequencies, where the aforementioned assumption might not be satisfied. 

\begin{remark}
	The parameter estimation performance of the proposed \ac{ML}-based algorithm, in particular range and velocity estimation, strictly depends on the dimension of the block of data sent, i.e., the product $N\cdot M$. Thus, the system parameters of Tab. \ref{tab:System-Parameters} can be easily tuned to achieve the desired levels of radar resolutions (modifying the bandwidth), acquisition time (based on the length of the \ac{OTFS} frame in time), maximum range, etc. Clearly, the \ac{CRLB} changes accordingly. Moreover, note that this is also possible thanks to \ac{OTFS} modulation, which is not sensitive to Doppler and delay effects.   
\end{remark}

\begin{remark}
	The (radar) range and velocity resolution in \eqref{eq:Radar-Resolution} indicates the minimum necessary targets spacing, in one of the two domain, such that both of them are distinguishable at the radar \ac{Rx}. This is not linked to the performance of our \ac{ML}-based detector, which is able to estimate the parameters accurately far beyond the resolution in \eqref{eq:Radar-Resolution}. Thus, there is a huge difference between targets identifiability and estimation performance.
\end{remark}

\begin{figure}
	\centering
	\includegraphics[scale=0.65]{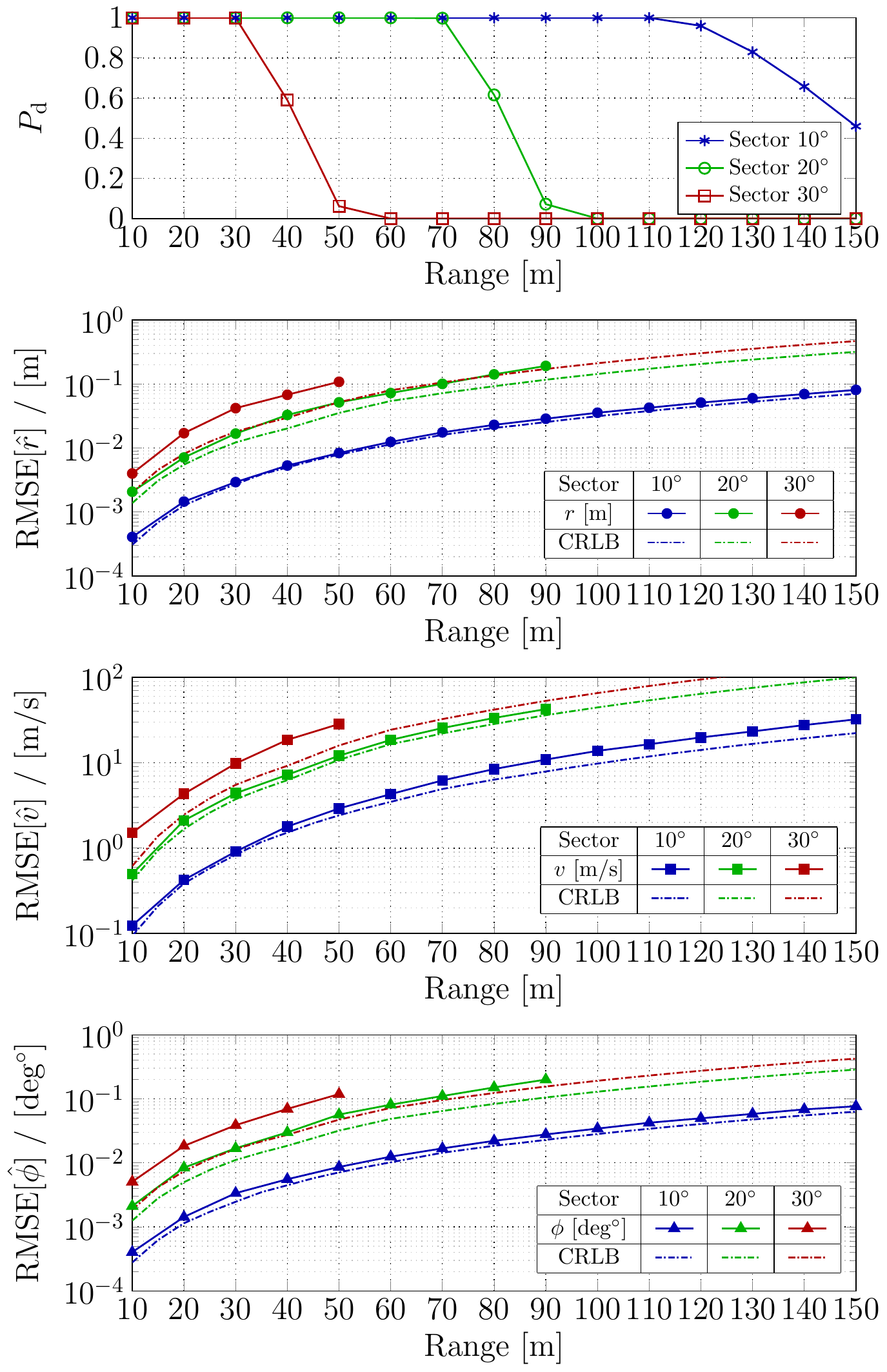}
	\caption{Detection phase. Single target at a different distance within the illuminated angular sector of specified coverage. RMSE performance with associated CRLB and detection performance. $N_a=128$.}
	\label{fig:Detection-Range-Na128}
\end{figure}

\begin{figure}
	\centering
	\includegraphics[scale=0.65]{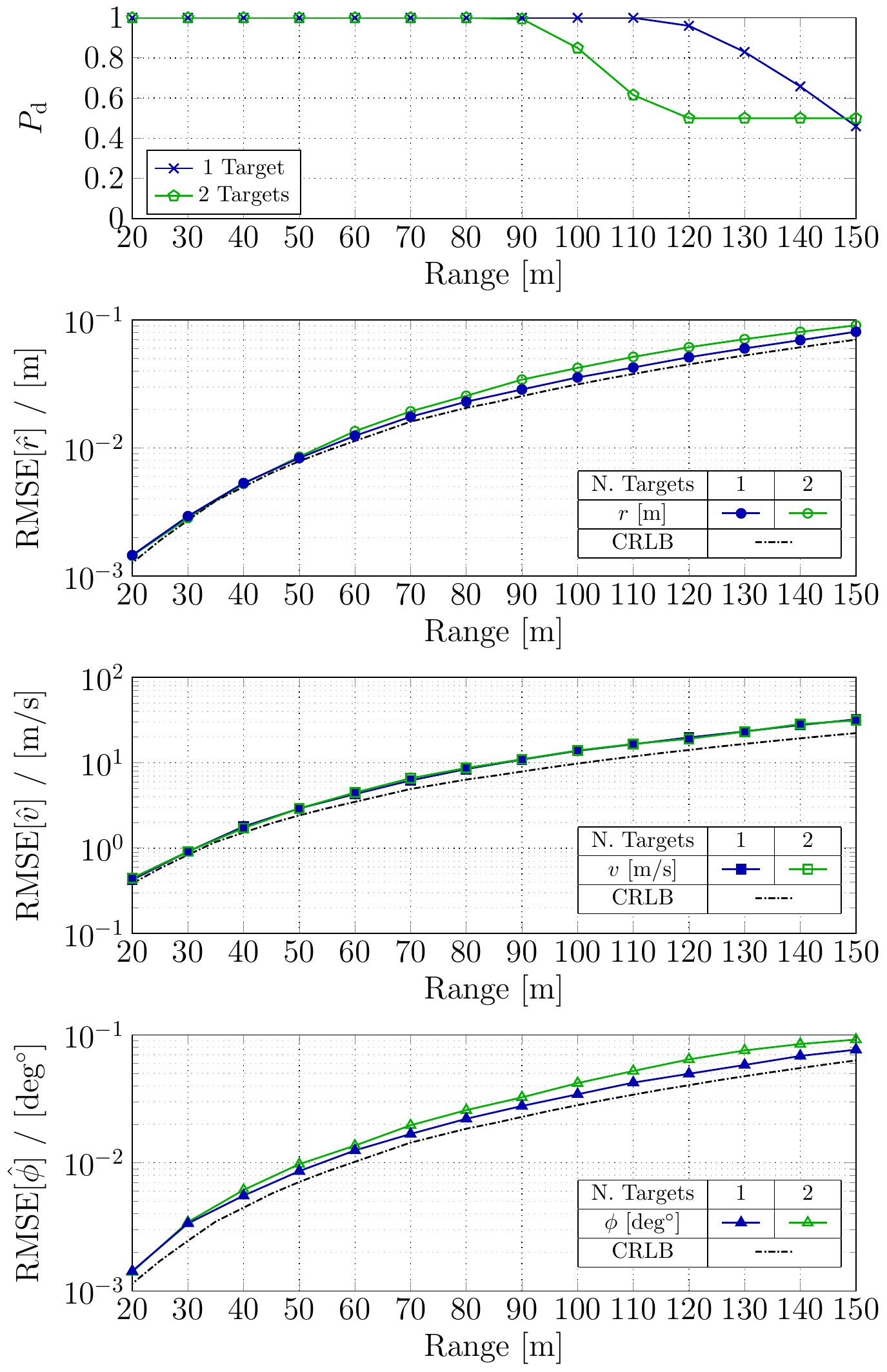}
	\caption{Detection phase. Two target, one located at $10$ [m] and the other moving at a different distance (x-axis) within the illuminated $10^\circ$ angular sector as shown in Fig. \ref{fig:Ex-Scenario-MovTarget}. RMSE performance with associated CRLB. Detection performance. $N_a=128$.}
	\label{fig:Detection-Range-Na128-2Targets}
\end{figure}

\begin{figure}
	\centering
	\includegraphics[scale=0.2]{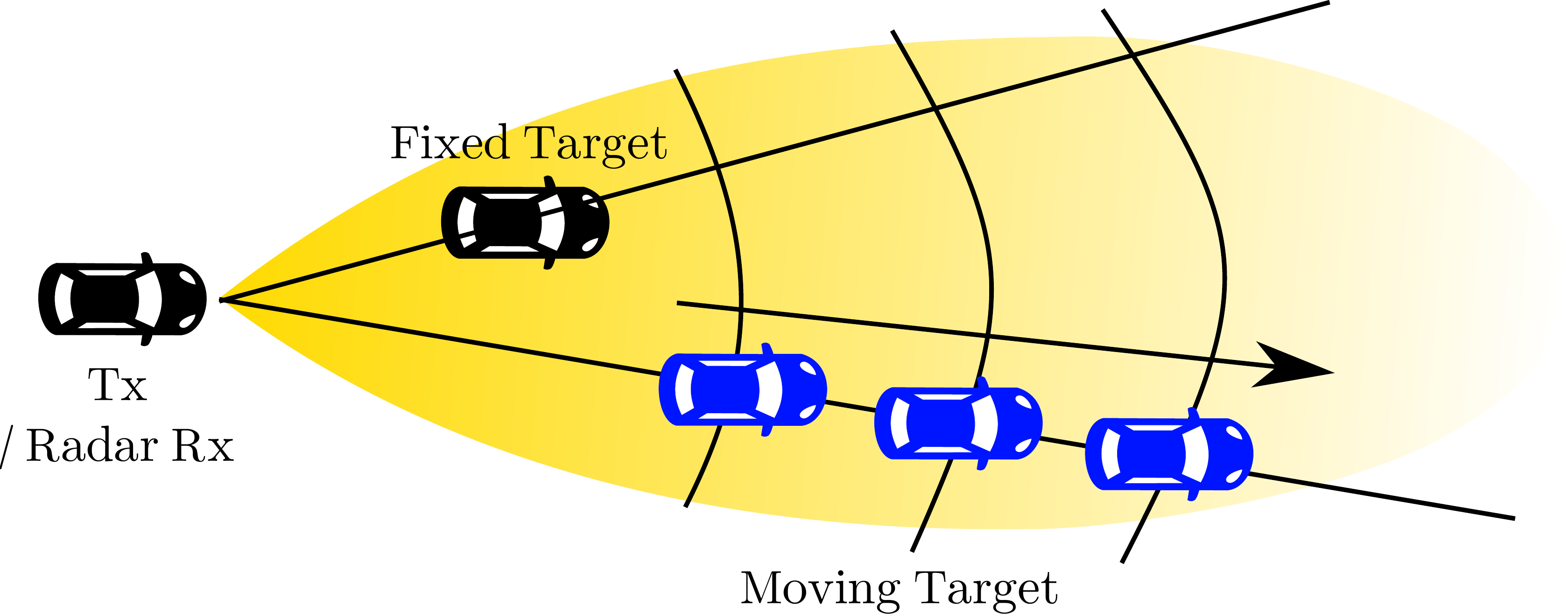}
	\caption{Example of scenario depicted in Fig. \ref{fig:Detection-Range-Na128-2Targets}. The fixed target (in black) masks the moving target, in blue, which changes its location within the illuminated angular sector.}
	\label{fig:Ex-Scenario-MovTarget}
\end{figure}

\begin{figure}
	\centering
	\includegraphics[scale=0.63]{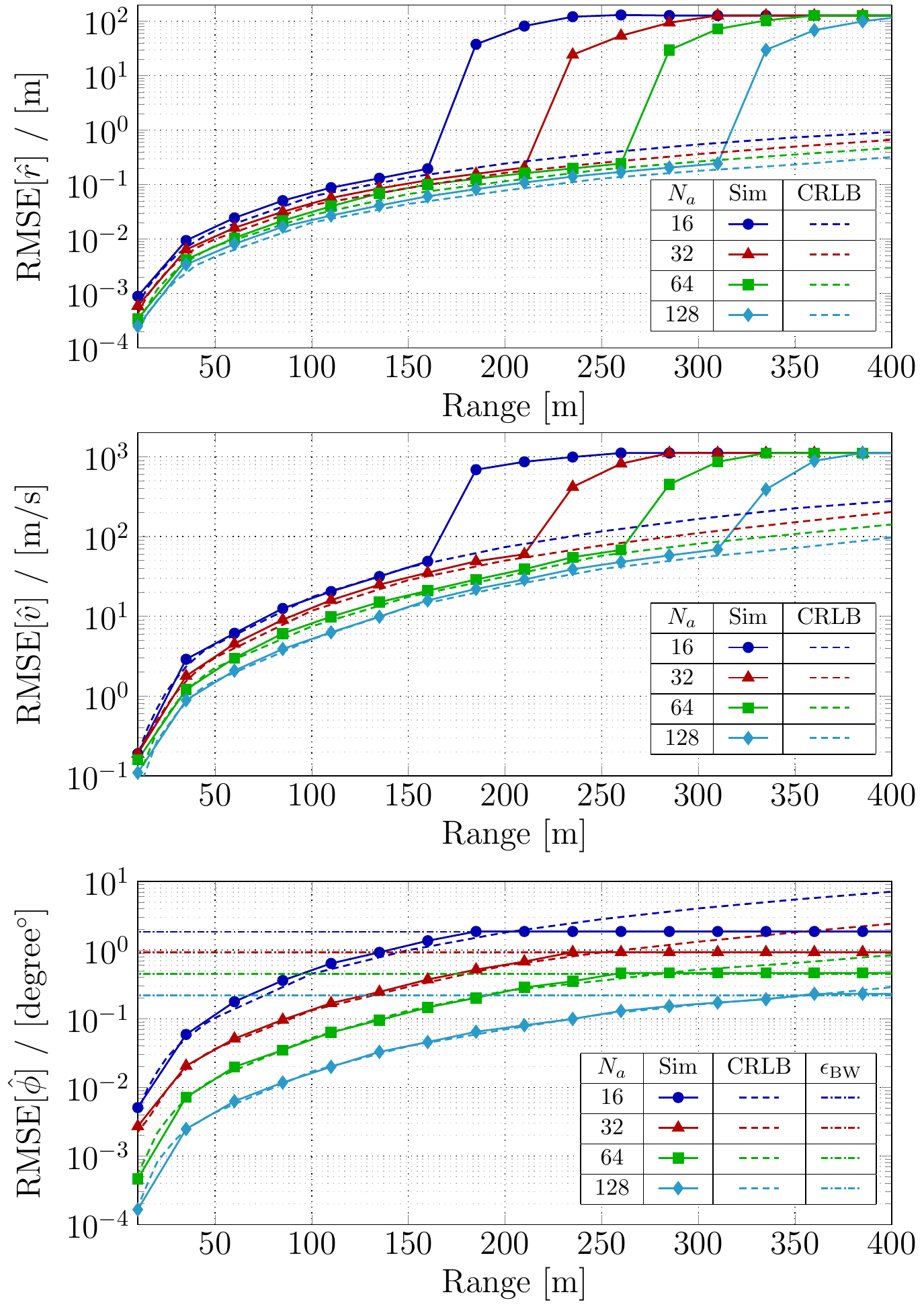}
	\caption{Tracking scenario. \ac{Tx} BF distinct beams towards three different targets. 
	}
	\label{fig:BF-Targets}
\end{figure}
\begin{figure}
	\centering
	\includegraphics[scale=0.2]{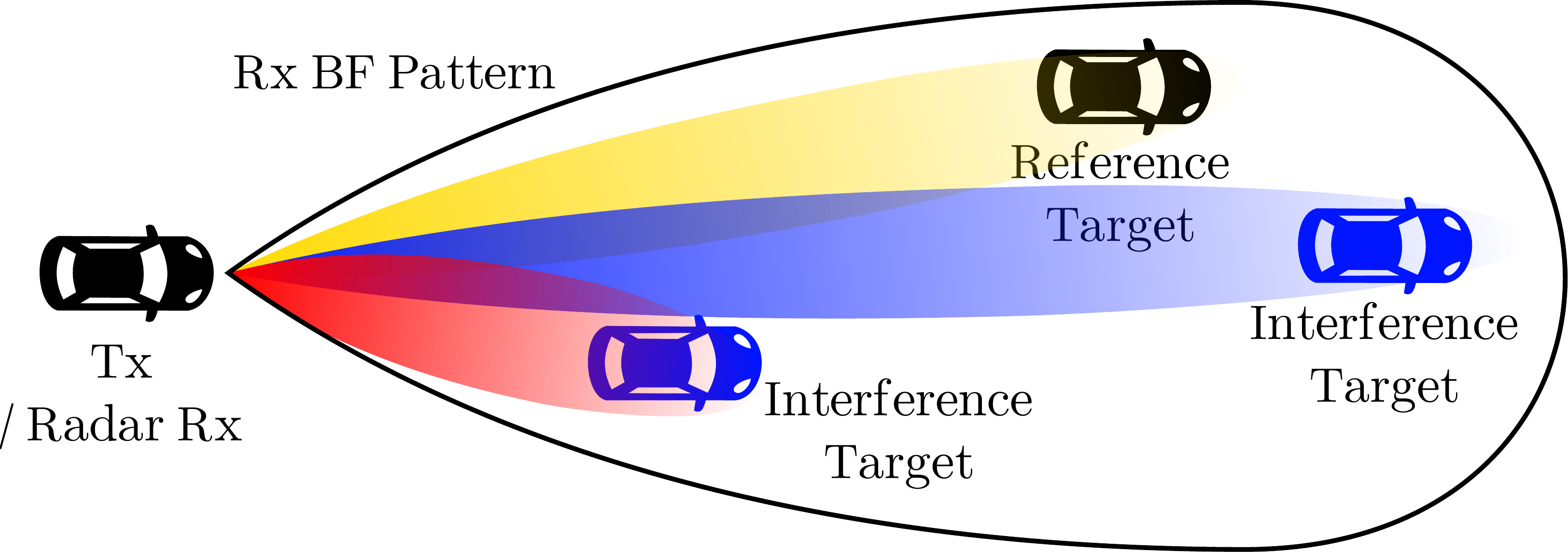}
	\caption{Example of scenario depicted in Fig. \ref{fig:BF-Targets}. The goal is to correctly estimate the parameters of the reference target (in black), while interference targets (in blue) lay within the same \ac{Rx} \ac{BF} pattern (depicted in Fig. \ref{fig:Rx-Beams} and with the black shape in this figure).}
	\label{fig:Ex-Scenario-BF-Targets}
\end{figure}

\subsection{Simulation Results}
Fig. \ref{fig:Detection-Range-Na128} shows the radar performance  in terms of \ac{PD}, range/velocity/\ac{AoA} estimation during the detection phase (Fig.\ref{fig:Operational-Modes-a}). When more than one target is considered within the simulation scenario, the \ac{PD} $P_{\rm d}$ is averaged w.r.t. all $P$ targets, i.e.,
\begin{equation}
	P_{\rm d}=\frac{\sum_{p=0}^{P-1}P_{\rm d}(p)}{P}\,,
\end{equation} 
where $P_{\rm d}(p)$ denoted the \ac{PD} of the $p$-th target.

First, note that, by considering an angular coverage of $10^\circ$ degrees (blue line), the maximum range to correctly identify a target, limited by the pathloss and thus different from the theoretical limit indicated in Table \ref{tab:System-Parameters}, is about $110$ m. For any distance between \ac{Tx} and target, the estimation performance of radar parameters of interest (range, velocity, and \ac{AoA}) follows the corresponding \ac{CRLB}. More in details, note that at the limit range of $110$ m, the \ac{RMSE} for range, velocity, and angle are respectively, $\simeq4\cdot10^{-2}$ [m], $\simeq1.6\cdot10^{1}$ [m/s], $\simeq4\cdot10^{-2}$ [degree$^\circ$]. As expected, given the system parameters, the velocity \ac{RMSE} is quite poor, while the others estimation performance are satisfactory. However, a proper \ac{BF} design towards targets, in a subsequent tracking phase, could improve the estimation performance maximizing the received \ac{SNR}, as showed in next results. As seen from Fig. \ref{fig:Detection-Range-Na128}, by increasing the angle sector from $10^\circ$ to $30^\circ$ degrees, the backscattered power gets smaller (less \ac{BF} gain), hence the maximum range significantly decreases. There exists a non-trivial tradeoff between the width of beams and radar performance. Wider angular sectors allow to explore the environment in less time, but with limited maximum range, while narrower sectors maximize the received power and the maximum target range, at the cost of a time consuming beam sweeping search. Clearly, \ac{RMSE} performance can not be computed if the \ac{PD} is equal to $0$, i.e., the target is not detected, thus \ac{RMSE} curves may stop at certain ranges, as visible in Fig. \ref{fig:Detection-Range-Na128}. 

Fig. \ref{fig:Detection-Range-Na128-2Targets} shows the performance of the \ac{SIC} technique presented in Algorithm \ref{alg:Joint-Det-Par-Est} during the detection phase in the scenario depicted in Fig. \ref{fig:Ex-Scenario-MovTarget}. Namely, the transmitter wishes to detect two targets, one at fixed distance of $10$ [m], the other with moving w.r.t. the x-axis, i.e., from $20$ to $150$ [m] (see Fig. \ref{fig:Ex-Scenario-MovTarget}). SIC is necessary because  the closer target (black car) will mask the further target (blue car) so that the latter cannot be detected. First, the first plot of Fig. \ref{fig:Detection-Range-Na128-2Targets}, referred to the \ac{PD}, shows that, when the moving target is located at ranges greater than $90$ [m], corresponding to the relative range beyond $80$ [m], the masking effect is not removed efficiently by the \ac{SIC} technique (the residual interference is remarkable), and the target at longer distance is not detected correctly. In fact, at the extreme point, the curve flats to $P_d=0.5$, because only one target out of two (clearly, the closet to the radar \ac{Rx}, i.e., the one fixed at $10$ [m]) is correctly detected. As clearly visible, the performance in terms of \ac{RMSE}, which considers in this case the estimation performance averaged w.r.t. the detected targets (note that the target located at 10 [m] is always detected correctly), slightly changes while considering one or two targets, as a confirmation of the effectiveness of the proposed algorithm. Note that the blue curves of Fig. \ref{fig:Detection-Range-Na128-2Targets} correspond to the blue ones of Fig. \ref{fig:Detection-Range-Na128}. 

Now we consider the tracking phase corresponding to Fig. \ref{fig:Operational-Modes-b}.The scenario takes into account one \ac{Tx} and three targets within an angular sector of $10^\circ$ degrees, as shown in Fig. \ref{fig:Ex-Scenario-BF-Targets}. Fig.\ref{fig:BF-Targets} shows the \ac{RMSE} performance of the reference target (black car) in the presence of other two targets (blue cars), 
Note that distance, velocity, and angular position of all three targets are randomly chosen at every Monte Carlo iteration, in such a way the complete masking effect presented in Fig. \ref{fig:Detection-Range-Na128-2Targets} does not occur, and an average of \ac{RMSE} results is finally computed. 
From Fig. \ref{fig:BF-Targets}, we observe that the \ac{RMSE} critically depends on the number of antennas. This is because the \ac{BF} gain grows proportionally with the number of antennas and increases the backscattered signal power. 
Moreover, note that a (reversed) ``waterfall'' behavior is shown for range and velocity estimation. This is because, even if the presence of the target is given for granted, low \ac{SNR} values might still lead to a large error during the Doppler-delay \ac{ML} maximization (see Algorithm \ref{alg:Joint-Det-Par-Est}). The waterfall behavior is typical of \ac{ML} estimators and has been extensively analyzed in \cite{gaudio2019effectiveness}. Also note that the \ac{AoA} \ac{RMSE} performance is upper limited by the 3-dB beamwidth of the beam pattern (see, e.g., \cite{friedlander2012transmit,visser2006array}). In fact, supposing that the target position lies within the 3-dB beamwidth, also the initial \ac{AoA} estimation (the upper and lower limit of matrix $\Omega$ in \eqref{eq:3D-Search}) is limited to that width. As a consequence, the \ac{RMSE} does not exceed a systematic error, indicated as $\epsilon_\mathrm{BW}$, calculated by averaging over random \ac{AoA} estimation realizations within the range of possibilities, i.e., between the upper and lower limit set by the 3-dB beamwidth of the beam pattern. 

\section{Conclusions}\label{sec:Conclusions}

In this paper,  we proposed an efficient \ac{ML}-based algorithm able to jointly perform target detection and radar parameters estimation, i.e., range, velocity, and \ac{AoA}, by using a \ac{MIMO} mono-static radar adopting on \ac{OTFS} modulation and operating in different modes. Simulation results demonstrate the robustness of the algorithm in term of both target identifiability and estimation by exploiting a \ac{SIC} mechanism. Interestingly, our proposed scheme is able to simultaneously send data streams between one to the number of RF chains, depending on different operational phases. There are a couple of interesting directions which are left as future works. These include the further optimization of the hybrid beamforming matrices, the comparison with other radar or/and communication waveforms, efficient target tracking method following some mobility models.

\section{Acknowledgment}
The work of Lorenzo Gaudio, Giuseppe Caire, and Giulio Colavolpe is supported by Fondazione Cariparma, under the TeachInParma Project. This research benefits from the HPC (High Performance Computing) facility of the University of Parma, Italy.

\bibliography{IEEEabrv,book}

\end{document}